\documentclass[twocolumn]{revtex4}

\usepackage{dcolumn}
\usepackage{bm}
\usepackage{amsmath}
\usepackage{amsfonts}
\usepackage{graphics}
\usepackage[dvips]{graphicx}
\usepackage{times}
\usepackage{setspace}
\usepackage{color}   
\usepackage{verbatim}
\usepackage[raggedright]{titlesec}
\usepackage[normalem]{ulem}
\usepackage{framed}
\usepackage{natbib}
\usepackage{hyperref}
\hypersetup{colorlinks=true,citecolor=blue, linkcolor=black,urlcolor=blue}
\usepackage{framed}
\usepackage{wrapfig}

\begin{document}
\title{Inequality and cumulative advantage in science careers: a case study of high-impact journals}
\author{Alexander M. Petersen}
\author{Orion Penner}
\affiliation{Laboratory for the Analysis of Complex Economic Systems and Laboratory of Innovation Management and Economics, IMT Lucca Institute for Advanced Studies, Lucca 55100, Italy}
 
\maketitle 
\begin{framed} 
\noindent 
For the long  published version, complete with full analysis, references, methods, and data summary, see:  \\
AM Petersen, O Penner (2014) {\it EPJ Data Science} {\bf 3}: 24. 
\href{http://www.epjdatascience.com/content/3/1/24}{DOI:10.1140/epjds/s13688-014-0024-y}\\ 
Send correspondence to:  \text{petersen.xander@gmail.com}
\end{framed}

Analyzing a large data set of publications drawn from the most competitive journals in the natural and social sciences we show that research careers exhibit the broad distributions of individual achievement characteristic of systems in which cumulative advantage plays a key role. 
 While most researchers are personally aware of the competition implicit in  the publication process, little is known about the levels of inequality at the researcher level. 
 
 Here we analyzed  both productivity  and impact measures for a large set of  researchers publishing in high-impact journals, accounting for censoring biases in the publication data by using distinct researcher cohorts defined over non-overlapping time periods. For each researcher cohort we calculated Gini inequality coefficients, with average Gini values  around 0.48 for total publications and 0.73 for total citations. For perspective, these observed values are well in excess of the  inequality levels observed for personal income in developing countries.

Investigating possible sources of this inequality,  we identify two potential mechanisms that act at the level of the individual that may play defining roles in the emergence of the broad productivity and impact distributions found in science. First, we show that the average time interval between a researcher's successive publications in top journals decreases with each subsequent publication. Second, after controlling for the time dependent features of citation distributions, we compare the citation impact  of subsequent publications within a researcher's publication record. We find that as researchers continue to publish in top journals, there is more likely to be a decreasing trend in the relative citation impact with each subsequent publication. This pattern highlights the difficulty of repeatedly producing research findings in the highest citation-impact echelon, as well as the role played by finite career and knowledge life-cycles. It also points to the intriguing  possibility of confirmation bias in the evaluation of science careers.

\begin{figure*}
\centering{\includegraphics[width=0.75\textwidth]{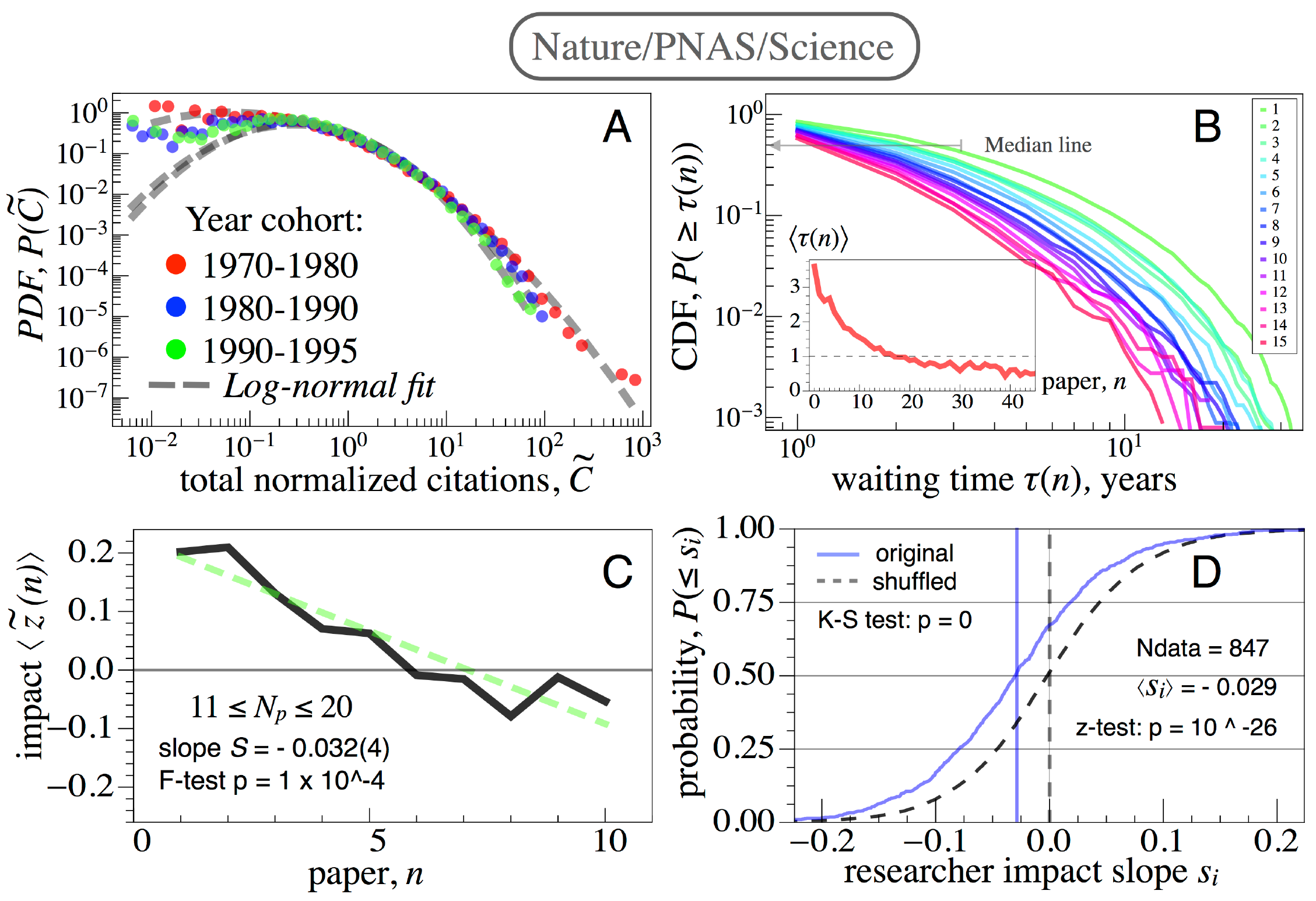}}
\caption{\label{fchampeconnat} Quantifying success in the Nature/PNAS/Science arena.  {\bf (A) Skewed citation distributions.} The citation measure $\tilde{C}^{j}_{i}$ is the total number of normalized citations (each paper is normalized to the average citation value of publications from the same year) a given researcher $i$ gained from papers in the journal set $j$. We account for censoring bias by aggregating researchers into non-overlapping subsets depending on when the researcher first published in the journal set. 
The  $P(\tilde{C})$ are extremely skewed, ranging over 4 orders of magnitude, and are well-fit by the log-normal distribution, except  for in the lower tail. {\bf (B) Increasing publication rate.} Shown are the complementary cumulative probability distributions, $P(\geq \tau(n))$, indicating the waiting time  $\tau(n)$ between two successive publications, for $n=1...15$.  By $n=10$ the observed likelihood of waiting  3  or more years,  $P(\geq 3 \vert n=10)$, falls to roughly $0.2$. (inset) The average waiting time, $\langle \tau^{j}(n) \rangle$, decreases significantly from 3.6 years for $n=1$ to  1 year by $n=16$. 
The values of $\langle \tau^{j}(1) \rangle = 3.6$ yrs.  Only research profiles with $L \geq 5$ years and $N_{p}\geq5$ are included. {\bf (C,D) Mean citation impact decreases with increasing $n$.} 
 For scientists with between 11 and 20 publications in the Nat./PNAS/Sci. journal set, we find a  significant negative trend  in  $\langle \tilde{z}(n) \rangle$ (black curve) with each successive publication. We also estimated the slope $s_{i}$ of individual $\tilde{z}_{i}(n)$ trajectories. The empirical cumulative  distribution $P(\leq s_{i})$  and the mean value $\langle s_{i} \rangle$ (vertical solid blue line) are shifted towards negative $s_{i}$ values. For comparison, we apply a shuffling technique to randomize $\tilde{z}_{i}(n)$ and then recalculate each $s_{i}$. The $P(\leq s_{i})$ for shuffled data (dashed black curve, mean indicated by vertical dashed gray line) are centered around 0. We apply the Kolmogorov-Smirnov test between the empirical and shuffled distributions, and the $p$-value confirms that the two sets of $s_{i}$ values belong to  different distributions.
In order to ensure that the relative citation impact $z_{p}$ of a given publication had sufficient time to stabilize within the journal set dataset, only publications published prior to 2002 for Nat./PNAS/Sci. were analyzed (since the publication citation counts  used were current as of our 2009 census year).  In order to reduce censoring bias arising form careers that started before the beginning of each data sample, we only included trajectories with the first publication year $y^{j}_{i,0}\geq 1970$.}
\end{figure*}

Our focal unit throughout the analysis is the scientific career, even though we use publication and citation counts as the central quantitative measure.
Our  data  comprises  412,498 publications drawn from 23 individual high-impact journals indexed by Thompson Reuters Web of Knowledge (TRWOK). From these data we extracted the publication trajectory of 258,626
individual scientists, where each trajectory is defined  {\it within  a  set of similar journals}. The three principal journal sets analyzed  are Nature/PNAS/Science, a collection of 14 high-impact economics journals, and a collection of  3 prestigious management science journals. For each analysis, we carefully selected comparable sets of researcher profiles using thresholds that controlled for possible censoring and cohort biases in the data. 

By analyzing researcher profiles within prestigious journals, we gather insights into the ascent of top scientists and the operational value of these highly-selective Òcompetitive arenas.Ó
Our analysis starts with the basic question: How do such skewed achievement distributions emerge, even within the highest-impact journals?  To this end, we used the longitudinal aspects of the data to quantify the role of  cumulative advantage in science careers, summarized in 3 parts: 

({\bf {\emph a}}) What are the levels of ``inequality'' within these high-impact distributions? 
For example, for researchers who had their first publication between 1970-1980, we calculated a Gini index $G=0.83$ (economics) and $G=0.74$ (Nat./PNAS/Sci.) and found that the top 1\% of researchers (comprised of  17 and 139 researchers, respectively) held a significantly disproportionate share of $26\%$  and $22\%$ of the total $\tilde{C}$ aggregated across all researchers in each distribution. For perspective, these inequality levels are in excess of those observed for personal income in developing countries.
Nevertheless, analysis of $G$ for different time periods indicates that both  productivity and impact equality is increasing  over time. 

({\bf {\emph b}})  How long does a researcher typically wait before his/her next high-profile publication? For each author, $i$, we define a sequence of waiting times, $\tau_{i}(n)$, for which the $n^{th}$ entry is the number of years between his/her publication $n$ and publication $n+1$ in a given journal set. The longest waiting time is typically between the first and second publication. For example, the average waiting time in both NEJM and Nat./PNAS/Sci. is roughly $\langle \tau(1) \rangle \approx 4$ years, whereas in the biology journal Cell and the physics journal PRL the initial mean waiting times are closer to  $\langle \tau(1) \rangle \approx 3$ years. With each successive publication, we found that $\langle \tau(n) \rangle$ decreases significantly, so that by the 10th publication the waiting time has decreased to roughly 1/2 of the initial waiting time $\tau(1)$. This shifting towards smaller waiting times with increasing $n$ is further evident in the entire distribution of waiting times, $P(\tau(n))$.

({\bf {\emph c}})  Focusing only a researcher's publications in selective high-impact journals, are a typical researcher's later publications more or less cited than their previous publications? To investigate the longitudinal variation in the citation impact, we map the citation count ${c}^{\ j}_{i,p,y}$ of the $n^{th}$ publication of researcher $i$, published in journal set $j$ to a $z$-score,
\begin{equation}
z_{i}(n) \equiv  \frac{\ln {c}^{\ j}_{i,p,y}(n) -   \langle \ln c^{j}_{y} \rangle}{ \sigma[ \ln c^{j}_{y}]} \ ,
\label{zn}
 \end{equation}
 which allows for comparison across time since publications are measured relative to publications from the same publication year $y$. 
In order to account for author-specific heterogeneity before we aggregate citation trajectories across scientists, we centered the $z$-score around
 the mean value $\langle z_{i} \rangle \equiv N_{p}^{-1} \sum_{n=1} z_{i}(n)$ calculated for the $N_{p}$ publications of a given scientist $i$. As a result, we obtain  the relative citation impact trajectory,
  \begin{equation}
  \tilde{z}_{i}(n) \equiv   z_{i}(n) - \langle z_{i} \rangle \  .
   \end{equation}
This normalization also helps in controlling for latent effects caused by disciplinary  variation  within the aggregated economies and multidisciplinary natural science journal sets,  which could affect the overall citation potential of a paper over time.
 Using these standardized $\tilde{z}_{i}(n) $ trajectories, we pooled the data across scientists, noting that  $\tilde{z}_{i}(n)$ is still measured in normalized units of the standard deviation $\sigma_{\ln c}$. For each journal set $j$ we observed a {\it negative trend} in $\langle \tilde{z}_{i}(n) \rangle$ (aggregate level) and a significant excess of negative trends in $\tilde{z}_{i}(n)$ (individual level),  see Fig.\ref{fchampeconnat} (C,D).
 
This result is indicative of  the complex  prestige system in science.  Finite career and knowledge lifecycles,  as well as the intriguing  possibility of identifying  institutional confirmation bias in the evaluation process of science careers, likely play a role in this the decreasing trend in $\tilde{z}_{i}(n)$. 
The latter explanation represents a possibly counterproductive role of cumulative advantage in science, since the publication of a high-impact publication early in the career, which may or may not be an appropriate predictor of sustainable impact in the future, nevertheless appears to facilitate additional future opportunities in these highly-competitive journals. 


\end{document}